\documentclass[10pt,twoside,a4paper,aps,twocolumn,preprintnumbers,nofootinbib,superscriptaddress]{revtex4-1}
\usepackage[utf8]{inputenc}
\usepackage[T1]{fontenc}
\usepackage[english]{babel}

\usepackage{hyperref}
\hypersetup{
    colorlinks=true,
    linkcolor=blue,
    filecolor=magenta,
    urlcolor=cyan,
}

\usepackage{verbatim}
\usepackage{url}
\usepackage{bbold}
\usepackage{latexsym}
\usepackage{lmodern} %Neat font
\usepackage{dcolumn}
\usepackage[pdftex]{graphicx}
\usepackage{booktabs}
\usepackage{blindtext}
\usepackage[includeheadfoot,margin=2cm]{geometry}
\usepackage{soul}

\usepackage{amssymb}

\usepackage{amsmath}
\usepackage{mathtools}
\usepackage{ae}
\usepackage{units}
\usepackage{icomma}
\usepackage{color}
\usepackage{enumerate}
\usepackage{epstopdf}
\usepackage{bbm}
\usepackage{cases}
\usepackage{lettrine}
\usepackage{rotating}
\usepackage{fancyhdr}
\usepackage{bm}
\usepackage{lastpage}
\usepackage[normalem]{ulem}
\usepackage{dsfont}
\usepackage{amssymb}
\IfFileExists{esvect.sty}{\usepackage{esvect}}{}
\usepackage[font=small,format=plain,labelfont=bf,textfont=it,indention=1cm,labelsep=period,justification=raggedright]{caption}
\usepackage{multirow}
\usepackage{array}
\usepackage{subfig}
\usepackage{cancel}
\usepackage{listings}
\usepackage{color}
\usepackage{textcomp}
\usepackage{tikz}
\usepackage{amsthm}
\usepackage{eso-pic}
\usepackage{mathrsfs}
\usepackage{numprint}
\usepackage[round-mode=places,locale=US]{siunitx}

\newcommand{\be}{\begin{equation}}
\newcommand{\ee}{\end{equation}}
\newcommand{\bea}{\begin{eqnarray}}
\newcommand{\eea}{\end{eqnarray}}

\newcommand{\CPT}{$\mathcal{CPT}$}
\newcommand{\Hbar}{$\overline{\mathrm{H}}$}
\newcommand{\nuHF}{$\nu_\mathrm{HF}$}

\newcommand{\nuHFlit}{$\nu_\mathrm{lit}$}

\begin{document}

%\linenumbers

\title{In-beam measurement of the hydrogen hyperfine splitting - towards antihydrogen spectroscopy}

\author{M.~Diermaier}
\affiliation{Stefan-Meyer-Institut für Subatomare Physik$,$ \"Osterreichische Akademie der Wissenschaften$,$ Wien 1090$,$ Austria}
\author{C.~B.~Jepsen}\altaffiliation{Present address: Department of Physics\\ Princeton University, New Jersey 08544, USA}
\affiliation{CERN$,$ Genève 1211$,$ Switzerland}
\author{B.~Kolbinger}
\affiliation{Stefan-Meyer-Institut für Subatomare Physik$,$ \"Osterreichische Akademie der Wissenschaften$,$ Wien 1090$,$ Austria}
\author{C.~Malbrunot}
\affiliation{CERN$,$ Genève 1211$,$ Switzerland}
\affiliation{Stefan-Meyer-Institut für Subatomare Physik$,$ \"Osterreichische Akademie der Wissenschaften$,$ Wien 1090$,$ Austria}
\author{O.~Massiczek}
\author{C.~Sauerzopf}
\author{M.~C.~Simon}
\author{J.~Zmeskal}
\author{E.~Widmann}
\email{eberhard.widmann@oeaw.ac.at}
\affiliation{Stefan-Meyer-Institut für Subatomare Physik$,$ \"Osterreichische Akademie der Wissenschaften$,$ Wien 1090$,$ Austria}

%\linenumbers

\setlength{\columnsep}{25pt}

\begin{abstract}

Antihydrogen, the lightest atom consisting purely of antimatter, is an ideal laboratory to study the  \CPT\  symmetry by comparison to hydrogen. %21
With respect to absolute precision, transitions within the ground-state hyperfine structure (GS-HFS) are most appealing by virtue of their small energy separation. %21 (1)
ASACUSA proposed employing a beam of cold antihydrogen atoms in a Rabi-type experiment to determine the GS-HFS in a field-free region. %21
Here we present a measurement of the zero-field hydrogen GS-HFS using the spectroscopy apparatus of ASACUSA's antihydrogen experiment. %18.
The measured value of \nuHF=$1~420~405~748.4(3.4)(1.6)$~Hz with a relative precision of $\Delta$\nuHF/\nuHF=$2.7\times10^{-9}$ constitutes the most precise determination of this quantity in a beam and verifies the developed spectroscopy methods for the antihydrogen HFS experiment to the ppb level. %38
Together with the recently presented observation of antihydrogen atoms 2.7~m downstream of the production region, the prerequisites for a measurement with antihydrogen are now available within the ASACUSA collaboration. 
%total=148

\end{abstract}

\maketitle

%\linenumbers

Investigations of the hydrogen atom have been a driving force for the discovery of more profound theories~\cite{KarshenboimEtAl2001} and contribute to the basis of physics through their prominent influence on the definition of fundamental constants~\cite{CODATA_2014}.
Most notable from a precision point of view are the recent measurement of the $1S-2S$ transition via two-photon spectroscopy~\cite{Parthey:2011ys} and the determination of the hyperfine splitting in hydrogen maser experiments in the early 1970s~\cite{Hellwig:1970,Karshenboim:2000,Essen:1971uq,Essen:1973,Ramsey:QED,RamseyRevMod90}.
The achieved absolute (relative) precisions are $10$~Hz ($4 \times 10^{-15}$) and  $2$~mHz  ($1.4 \times 10^{-12}$), respectively.
A revival of the interest in hydrogen is founded on prospects of antihydrogen (\Hbar) research~\cite{CharltonEtAl1994,Holzscheiter:2004ur}.
The structure of the simplest antiatom, consisting of a positron bound to an antiproton is predicted to be identical to that of hydrogen, if the combined symmetry of charge conjugation, parity, and time reversal (\CPT) is conserved.
Hence, antihydrogen spectroscopy promises precise tests of the \CPT\ symmetry, which is a cornerstone of the Standard Model of particle physics.
A vivid physics program is currently underway at the Antiproton Decelerator of CERN aiming at spectroscopic~\cite{HoriWalzAD,Andresen:2010jba,Enomoto:2010uq,GabrielseEtAl2012,Kuroda:2014fk} and gravity tests~\cite{PerezEtAl2012,AghionEtAl2014} along with other \CPT\ tests like the neutrality of antihydrogen~\cite{Amole:2014,AhmadiEtAl2016} as well as measurements of the charge-to-mass ratio~\cite{Ulmer:2015} and magnetic moment~\cite{DiSciacca:2013wd} of the antiproton.

Among the spectroscopic tests of \CPT, the comparison of the GS-HFS of hydrogen and antihydrogen has the potential to reach the highest sensitivity on an absolute energy scale~\cite{Bluhm:1999vq,KosteleckyVargas2015,Widmann:2013qy}.
However, the aforementioned most precise measurement of this quantity for hydrogen was made using a maser~\cite{Ramsey:QED}.
Such a technique is not applicable to antimatter which would annihilate with the confining matter enclosure.
The measurement proposed by the ASACUSA collaboration at the Antiproton Decelerator of CERN therefore makes use of a beam of cold antihydrogen atoms~\cite{Mohri:2003wu,Widmann:2001fk}.
In addition to avoiding wall interaction, the actual measurement takes place in a field-free region, ultimately allowing for higher precision compared to the observation of resonant quantum transitions between the hyperfine states in trapped antihydrogen in a high-field environment~\cite{Amole:2012bh}.

\begin{figure}[htbp]
\begin{center}
\includegraphics[width=1\columnwidth]{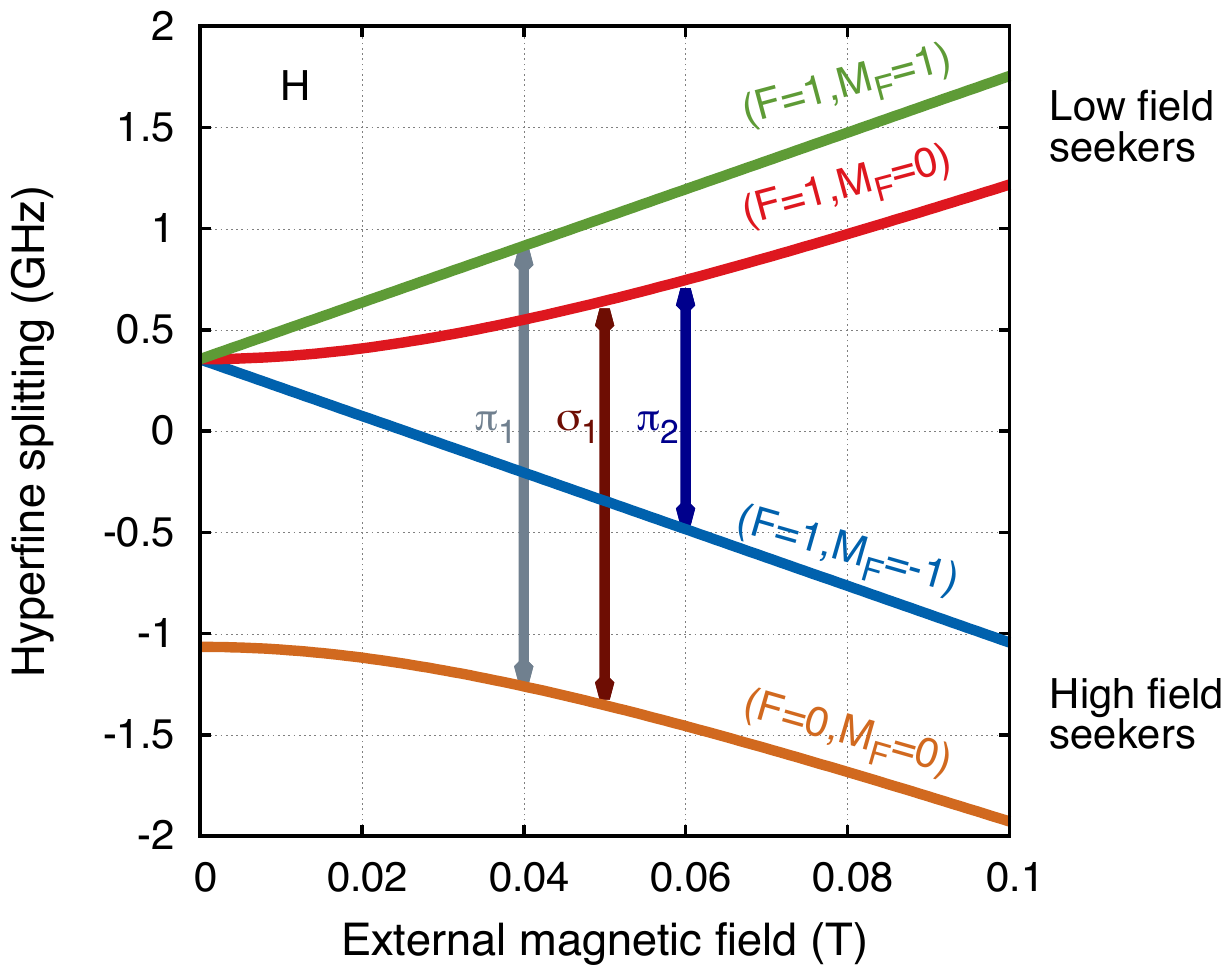}
\caption{
%xxx\internallinenumbers
\textbf{Ground state hyperfine splitting in hydrogen.}
The Breit-Rabi diagram shows the energy levels in ground-state hydrogen as a function of the strength of an external magnetic field. The 4 hyperfine states separate into a singlet state and a triplet state, which exhibit different Zeeman shifts. The states with a positive or negative slope are named low or high field seekers, respectively. Three possible hyperfine transitions between \textit{lfs} and \textit{hfs} are denoted by arrows, the $\sigma_1$-transition occurs between the states ($F$=$1$,$M_F$=$0$) and ($F$=$0$,$M_F$=$0$).}
\label{fig:BreitRabi}
\end{center}
\end{figure}

Rabi-type magnetic resonance spectroscopy~\cite{Rabi:1938tq,Rabi:1939tl} applies rotating (or oscillating) magnetic fields to induce quantum transitions and exploits the force of magnetic field gradients on the state-dependent magnetic moment of atoms (or molecules) in order to spatially separate the atoms in a beam with respect to their quantum states (Stern-Gerlach separation).
Typically magnetic sextupole fields are employed to focus atoms in low-field-seeking states (\textit{lfs}) and defocus high-field-seekers (\textit{hfs}).
In the case of ground-state hydrogen the hyperfine structure consists of a lower lying singlet state with total angular momentum quantum number $F$=$0$ ($\mathbf{F}$=$\mathbf{S}^p$+$ \mathbf{S}^e$ with $\mathbf{S}^p$ and $\mathbf{S}^e$ being proton and electron spin, respectively) and a triplet state $F$=$1$.
As illustrated by the Breit-Rabi diagram in Fig.~\ref{fig:BreitRabi} the triplet state degeneracy is lifted in the presence of a magnetic field.
The singlet state and the triplet state with magnetic quantum number $M_\mathrm{F}$=$-1$ are \textit{hfs}, while the other two states ($F$=$1$,~$M_\mathrm{F}$=$0$,$1$) are \textit{lfs}.
In the present experiment the $\sigma_1$-transition from ($F$=1,~$M_\mathrm{F}$=$0$) to ($F$=$0$,~$M_\mathrm{F}$=$0$) has been studied~\cite{ramsey1986molecular,Nafe:1948}.
The Zeeman-shifted frequencies at various external magnetic field strengths were determined for subsequent extraction of the zero-field value and resulted in

\begin{equation}
\nu_\mathrm{HF} = \unit[1~420~405~748.4(3.4)(1.6)]{Hz}.
 \end{equation}
 
The numbers in brackets are the one standard deviation (1$\sigma$) statistical and systematic uncertainties.
Added in quadrature the total uncertainty of  \unit[3.8]{Hz} constitutes an improvement by more than an order of magnitude in comparison to the previously achieved best precision by Rabi-type spectroscopy of \unit[50]{Hz}~\cite{Prodell:52,Kusch:1955aa}.
Our result is in agreement within one standard deviation with the literature value of \mbox{\nuHFlit =\unit[1~420~405~751.768 (2)]{Hz}}, which relies on the more precise hydrogen maser measurements~\cite{Hellwig:1970} (see discussion in \S6 of~\cite{Karshenboim:2000}).

\section{Results}

\subsection{Experimental Setup}

\begin{figure*}[htbp]
\begin{center}
\includegraphics[width=1\textwidth]{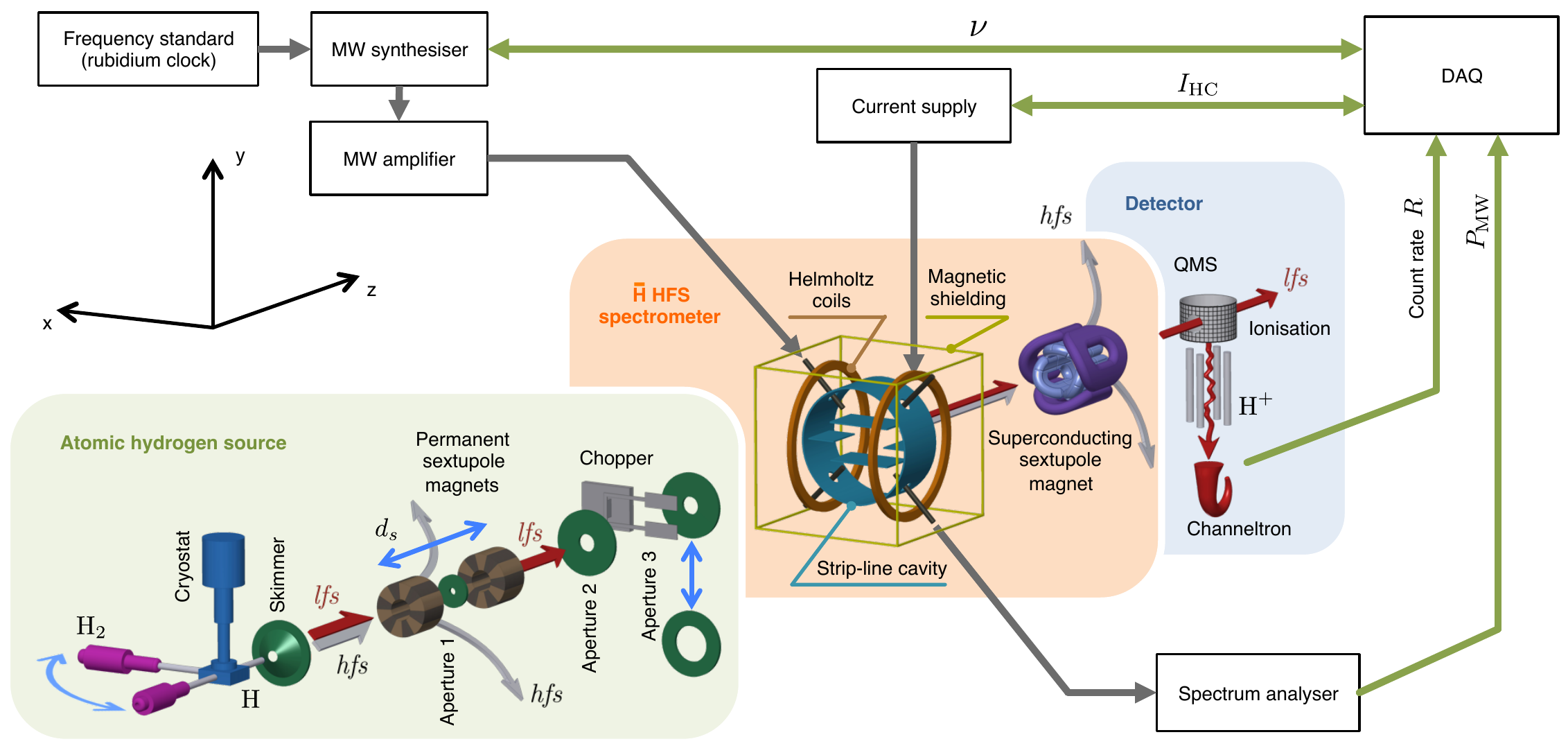}
\caption{
%xxx\internallinenumbers
\textbf{Atomic hydrogen beam setup.}
Illustration of the three main components of the Rabi-type experimental setup (not to scale). Green panel: the source of cold, polarised, and modulated atomic hydrogen. Orange panel: the hyperfine spectrometer of ASACUSA's antihydrogen experiment. Blue panel: the detector. The source consists of a microwave driven plasma for dissociation of H$_2$, a cryostat for cooling the atomic hydrogen beam in a PTFE tubing, two permanent sextupole magnets for polarisation and velocity selection, and a tuning fork chopper for beam modulation. The hyperfine spectrometer consists of a state-conversion cavity of strip-line geometry and Helmholtz coils enclosed in a cuboidal Mu-metal shielding followed by a superconducting sextupole magnet for spin-state analysis. The detector employs a quadrupole mass spectrometer for selective mass=$1$ ion (H$^+$) counting after ionisation. The count rate is acquired as a function of the driving frequency supplied to the cavity. 
}
\label{fig:Hbeam}
\end{center}
\end{figure*}

The main components of the experiment are a source of cooled and polarized atomic hydrogen, the hyperfine spectrometer of the \Hbar\ hyperfine splitting (HFS) setup (i.e. a microwave cavity and a superconducting sextupole magnet), and a hydrogen detector (cf. Fig.~\ref{fig:Hbeam}).
The atomic hydrogen source maintains a microwave driven plasma in a pyrex cylinder to dissociate molecular hydrogen ($\text{H}_2 \rightarrow \text{H} + \text{H}$)~\cite{McCullough}.
Hydrogen atoms are allowed into the first vacuum chamber through a PTFE tubing, which is kept under cryogenic temperatures in order to cool the hydrogen atoms and hence reduce their velocity~\cite{Walraven}.
Two tubing configurations are used in which the plasma-containing pyrex cylinder is either mounted perpendicular to or on axis with the beam.
In the first case, a 90$^\circ$ bent tubing assures an efficient and complete interaction of the hydrogen atoms with the cold PTFE surface.
In the latter case, a straight tubing keeps the recombinations caused by wall interactions down to a minimum.
The cooled atomic hydrogen beam is directed onto a skimmer of \unit[1]{mm} in diameter and reaches the second, differentially-pumped chamber, which houses two permanent sextupole magnets with a pole field of $\sim$\unit[1.3]{T} at a radius of \unit[5]{mm} over a mechanical length of \unit[65]{mm} each~\cite{Thonet:2016}.
In addition to providing the initial spin-polarisation, those sextupole magnets are moveable and feature a midway aperture (aperture 1) to allow for the adjustable selection of a narrow velocity range.
As the focusing length depends on the beam velocity, only a certain velocity component is focused onto the aperture and can pass, while the off-axis portions of all other components are blocked.
The variable distance to the aperture located at half the distance between the sextuple magnets ($d_\text{s}$) therefore selects a velocity component.
The resulting velocity distribution is much narrower than a Maxwell-Boltzmann distribution and roughly of Gaussian shape.
%%%
The spin-polarised and velocity-selected hydrogen beam passes another aperture (aperture 2) and is then modulated by a tuning fork chopper in the next differentially-pumped section.
The modulation adds time-of-flight measurements to the beam diagnostic tools as well as suppression of background originating from residual hydrogen via lock-in amplification.
Downstream of the chopper, apertures of different diameters (aperture 3) can be installed in order to produce different beam sizes at the entrance of the microwave cavity.

The \Hbar\ HFS spectrometer has been designed with an open diameter of \unit[100]{mm} since a large acceptance is crucial in view of small \Hbar\ production rates.
The amplitude of the oscillating magnetic field $B_{\mathrm{osc}}$ has to be sufficiently uniform over the large open diameter in order to guarantee a trajectory-independent state-conversion probability.
This requirement is best met by a cavity of so-called strip-line geometry~\cite{Kroyer:1098371,Silkethesis}.
Two highly transparent meshes confine the microwaves at the entrance and exit of the state-conversion cavity, which are separated by half a wavelength of the hyperfine splitting transition ($L_{\mathrm{cav}} \sim \lambda_{HF}/2  \sim$ \unit[105.5]{mm}).
A standing wave forms between them and as a consequence $B_{\mathrm{osc}}$ is not constant along the beam propagation direction, causing a \emph{double-dip} resonance line-shape.
The origin of this structure is outlined below and explained in detail in the Methods.
The cavity length and the beam velocity $v_H$ define the interaction time of the hydrogen atoms with the microwave field $T_{\mathrm{int}}=L_{\mathrm{cav}}/v_H$ and restrict the achievable resonance line width to $\sim T_{\mathrm{int}}^{-1}$.
A synthesiser coupled to an external rubidium clock for frequency stabilisation produces microwaves, which are fed radially to the cavity via an antenna after amplification.
On the opposite side of the cavity another antenna is used for pick-up and monitoring of the microwave power ($P_{\text{MW}} \propto B_{\mathrm{osc}}^2$) using a spectrum analyser.
Helmholtz coils are mounted onto the cavity to generate a homogeneous external magnetostatic field $B_\mathrm{stat}$, parallel to $B_\mathrm{osc}$, and of several Gauss in magnitude at the interaction region for fine control of the Zeeman splitting.
A current source with a relative stability of \unit[20]{ppm} supplies the Helmholtz coils' current $I_\mathrm{HC}$, which is independently monitored by an amperemeter.
$I_\mathrm{HC}$ is directly proportional to $B_\mathrm{stat}$ and turned out to be a better proxy for the magnetic field inside the cavity than a dedicated external magnetic field measurement.
The microwave cavity and the Helmholtz coils are surrounded by a two-layer cuboidal Mu-metal shielding to block the Earth's magnetic field as well as the fringe field of the closely succeeding superconducting sextupole magnet.
Owing to the pole strength of up to $\unit[3.5]{T}$, this magnet generates sizeable magnetic field gradients despite the large open diameter of \unit[100]{mm}. The integrated gradient amounts to \unit[150]{T/m} and ensures refocusing of \unit[50]{K} \textit{lfs} atoms within a distance of $\sim$\unit[1]{m}.

The detection of hydrogen suffers from a large background rate and small efficiencies.
A crossed-beam quadrupole mass spectrometer (QMS) with a \unit[3]{mm} opening ionises beam atoms and residual gas by electron impact and selectively guides protons to a channeltron for efficient single mass=$1$ ion counting.
The QMS can be moved two-dimensionally in the plane perpendicular to the beam for optimising count rates and investigating beam profiles.
Ultrahigh vacuum conditions ($p \leq$ \unit[5$\times$10$^{-10}$]{mbar}), achieved by combining two-stage turbo-molecular pumping and non-evaporable-getter pumps, lead to count rates of tens of kHz at the QMS for a typical $\text{H}_2$-flowrate of \unit[1.8$\times$10$^{17}$]{s$^{-1}$}.

\subsection{Measurement procedure}

\begin{figure}[htbp]
\includegraphics[width=0.87\columnwidth]{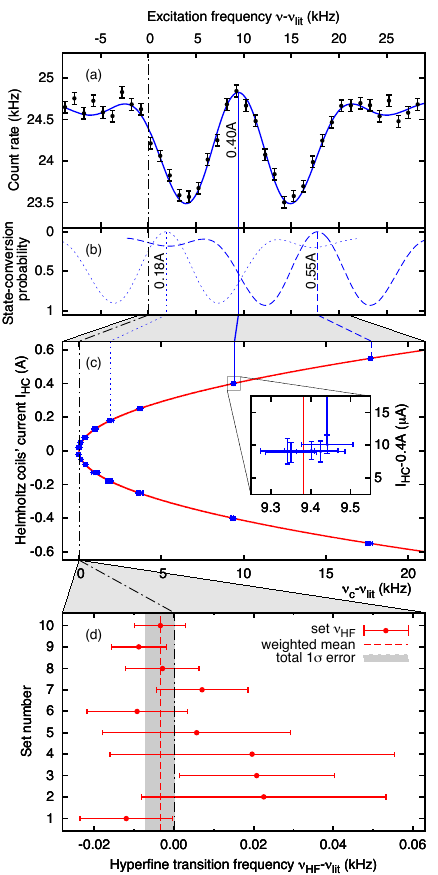}
\caption{
%%xxx\internallinenumbers
\textbf{Resonance spectrum and zero-field value extraction.}
(a) Data of one \emph{cycle} of \emph{set} 8 at a Helmholtz coils' current $I_\text{HC}$=\unit[400]{mA} fitted with the resonance curve ($\mathcal{F}_R$, full blue line, see equation~(\ref{eq:lineshapeR}) in Methods) to extract the \emph{central frequency} $\nu_c$, which is Zeeman-shifted to \mbox{values >~\nuHF .}
A dashed-dotted black line at 0 is drawn through all the plots to represent \nuHFlit .
(b) State-conversion probabilities ($\mathcal{F}$) as obtained from the fit ($\mathcal{F}_R$) of two other \emph{cycles} of the same \emph{set}, but at different settings of $I_\mathrm{HC}$ (dotted blue line \unit[180]{mA}, dashed blue line \unit[550]{mA}, data omitted for clarity).
(c) $\nu_c$ of all 80 \emph{cycles} of \emph{set} 8 (16 different values of $I_\mathrm{HC}$, 5 \emph{cycles} each) plotted against $I_\mathrm{HC}$ for extraction of the zero-field hyperfine splitting \nuHF\ using the Breit-Rabi fit function $\nu_\sigma'$ (red line) of equation~(\ref{eq:Rabiformulafit}). The inset is a zoom into the group of 5 \emph{cycles} at  $I_\text{HC}$=\unit[400]{mA} illustrating the typical size of the frequency and current standard deviations of each data point.
(d) The resulting \nuHF\ as deviation from \nuHFlit\ for the 10 \emph{sets} (red error bars) and their weighted mean value (dashed red line) with the $1\sigma$ total uncertainty as grey-shaded area.
}
\label{fig:big_plot}
%\end{center}
\end{figure}
%\end{figure*}

The dissociation plasma was operated under stable standard conditions.
Before starting frequency scans the microwave power $P_\mathrm{MW}$ supplied to the cavity was adjusted to yield the largest state-conversion probability by observing a Rabi oscillation.
A single measurement \emph{cycle} was obtained by scanning the frequency once in a random sequence across the desired range.
Typically, this included 39 frequency points distributed over $\sim$\unit[40]{kHz}.
At each frequency point the channeltron events of the QMS were summed several times for typical intervals of \unit[5-60]{s} from which an average count rate
was retrieved.
Such \emph{cycles} over the frequency range were repeated on average 5 times with changing random sequences to result in a complete \emph{scan} at a given $I_\mathrm{HC}$.
This was repeated at different values and polarity of $I_\mathrm{HC}$ to yield a \emph{set} of \emph{scans} suitable for determination of the field-free hyperfine splitting.
The number of $I_\mathrm{HC}$ values per \emph{set} ranged from 6 to 16.
In total 10 such \emph{sets} have been recorded, which differ in various of the experimental settings and arrangements (cf. table~\ref{tab:DAQasdf}).

\subsection{Raw data corrections}

Initially a fit as described below was applied to the detected count rates. 
Two systematic effects were identified in the residuals and corrected for.
The first correction compensates slow time drifts.
The second correction concerns a type of \emph{memory-effect}, which became evident in an increased likelihood of observing positive or negative residuals if the previous data point was taken at higher or lower count rate, respectively.
This indicated, that the settling of the hydrogen rate in the detection chamber following a change of the excitation frequency had a non-negligible time constant when compared to measurement time at each frequency step.
These two effects were corrected for at the raw data level and led to an improvement of the fit quality without affecting the extracted \nuHF\ values.
The application of a random sequence of frequencies in the \emph{cycles} seemed to suppress systematic impacts of the drift and the memory-effect below the statistical sensitivity.

\subsection{Analysis}

The \emph{central frequency} $\nu_c$ was extracted from every \emph{cycle} by a fit to the spectrum as illustrated in Fig.~\ref{fig:big_plot}a, where the excitation frequency $\nu$ is given as the difference to \nuHFlit .
The \emph{double-dip} line shape originates from the sinusoidal dependence of $B_{\mathrm{osc}}$ along the beam axis, which follows half a cosine period. At the actual transition frequency the highest count rate between the two dips is observed.
The theoretical line shape for a mono-energetic beam is well understood and accurately described within the framework of the two-level system with the interaction Hamiltonian

\begin{equation}
\mathcal{\hat{H}}_{\mathrm{int}}= - \hat{\pmb{\mu}} \cdot \mathbf{B}_{\text{osc}}(t) ,
\end{equation}

where $\hat{\pmb{\mu}} $ is the magnetic moment operator as defined in equation~(\ref{eq:MagMomentOperator}).
The time dependence of the magnetic field includes a $\cos(\pi t / T_{\text{int}})$-term in addition to the microwave oscillations.
The resulting equations were solved numerically to obtain the state-conversion probability as a function of the frequency $\nu$ and amplitude $B_{\mathrm{osc}}$ of the driving field for a mono-energetic beam. 
A realistic fit function $\mathcal{F}$ for the measured state-conversion probabilities was obtained by convolution of the shape for mono-energetic beams with a velocity distribution as described in the Methods.
Consequently, the fit function could extract the physical parameters $B_{\mathrm{osc}}$, the mean velocity of the polarised atomic hydrogen beam $\bar{v}_H$, and the width of the velocity distribution $\sigma_v$ in addition to $\nu_c$ of the transition.
Two further fit parameters of less relevant physical content scaled the state-conversion probability to the count rate and correspond to the count rate baseline $R_0$ and the count rate drop $\Delta R$ for complete state conversion.
In the final analysis only $\nu_c$ was extracted from every \emph{cycle} individually.
For  $B_{\text{osc}}$ a relation to the monitored microwave power was established based on the complete available data.
This enabled individual fixation of this parameter for every \emph{set} and avoided non-converging fits due to a strong correlation of $B_{\text{osc}}$ with $\Delta R$.
For $\bar{v}_H$ and $\sigma_v$ a common fit value for a complete \emph{set} was used, as all settings of direct impact on the beam velocity remained unchanged during data collection of a \emph{set}.

As illustrated in Fig.~\ref{fig:big_plot}a the line shape thus obtained resulted in good fits to the observed count rates at all $I_\mathrm{HC}$ settings with reduced $\chi^2$ values close to unity as summarized in table~\ref{tab:DAQasdf}.
The reliability of the fit function was important since $\nu_c$ could be extracted with typical statistical uncertainties on the order of tens of Hz while the width of the \emph{double-dip} structure is on the order of tens of kHz.
In Fig.~\ref{fig:big_plot}c, the extracted $\nu_c$ value of each \textit{cycle} of \emph{set} 8 are plotted against the Helmholtz coils' current $I_{\text{HC}}$ at which it was recorded.
The Zeeman-shifted frequency of the $\sigma_1$-transition $\nu_\sigma (B_\text{stat})$ has only a second order dependence on the static external magnetic field  $B_\mathrm{stat}$ as apparent from the Breit-Rabi diagram (Fig.~\ref{fig:BreitRabi}) and described by the Breit-Rabi formula~\cite{BreitRabi:1931}

\begin{equation}
\label{eq:Rabiformula}
\begin{split}
\nu_\sigma (B_\text{stat})= \sqrt{\nu_\mathrm{HF}^2+\left( \frac{\mu_{+} }{h} \right)^2 B_{\text{stat}}^2} ,\\
\mu_{+} = |g_e| \mu_B + g_p \mu_N ,
\end{split}
\end{equation}

with
\mbox{$\mu_B$=\unit[5.788~381~8012$\times$10$^{-5}$]{eVT$^{-1}$}} and
\mbox{$\mu_N$=\unit[3.152~451~2550$\times$10$^{-8}$]{eVT$^{-1}$}}
being respectively the Bohr and nuclear magneton,
\mbox{$g_e$=-$2.002~319~304~361~82$} and
\mbox{$g_p$=$5.585~694~702$}~\cite{CODATA_2014}
being respectively the $g$-factors of the electron and proton,
and $h$=$2\pi\hbar$ the Planck constant.
In order to extract the zero-field hyperfine transition frequency \nuHF\ a fit function $\nu_{\sigma}'$ was required, that used $I_\text{HC}$ as a variable.
A factor $k$ converting $I_{\text{HC}}$ to a magnetic field and a residual field $B_{\text{res}}$ at $I_{\text{HC}}=0$ added two further fit parameters and established a linear relation to $B_\text{stat}$, which enters the Breit-Rabi formula

\begin{equation}
\label{eq:Rabiformulafit}
\begin{split}
\nu_{\sigma}'(I_{\text{HC}};\nu_\mathrm{HF},k,B_{res}) = \\
= \sqrt{\nu_\mathrm{HF}^2+\left( \frac{\mu_{+}}{h} \right)^2 \left( k I_{\text{HC}}+ B_{\text{res}} \right)^2} .
\end{split}
\end{equation}

The notation for the fit function separates the variable from the parameters by a semicolon.
The zero-field values \nuHF\ as obtained via this Breit-Rabi fit are plotted in Fig.~\ref{fig:big_plot}d as the deviation from \nuHFlit .

\subsection{Systematic tests}

The following experimental arrangements and conditions have undergone changes for the 10 \emph{sets} (summarised in the top part of table~\ref{tab:DAQasdf}).
The beam velocity varied due to different settings of  $d_\text{s}$ and the temperature of the PTFE tubing.
The first three \emph{sets} operated with the straight PTFE tubing then the bent tubing was used.
The need for an improved monitoring of $I_\text{HC}$ and the advantage of a faster data acquisition scheme based on the total count rate instead of the lock-in amplifier signal became evident in a preliminary evaluation of the first three \emph{sets} and motivated the additional changes at that stage.
Two opening diameters for aperture 3, resulting in different beam sizes at the entrance of the cavity, were also investigated.
This is of special interest as an even larger beam diameter is expected for the  \Hbar\ HFS spectroscopy.
Additionally, the last 4 \emph{sets} were performed with a second cavity of the same but slightly upgraded design.
Three aspects were only changed for individual \emph{sets}.
For \emph{set} 7 only one instead of two layers of magnetic shielding were used, for \emph{set} 3 the superconducting sextupole worked with a larger magnetic field strength leading to a shorter focal length, and for \emph{set} 4 the direction of the static magnetic field (Helmholtz coils) was not reversed.

The obtained results for \nuHF\ of the 10 \emph{sets} by firstly fitting all \emph{cycles} in a \emph{set} using the fit function~(\ref{eq:lineshape}) and secondly the Breit-Rabi fit~(\ref{eq:Rabiformulafit}) are presented in Fig.~\ref{fig:big_plot}d.
Additionally the average reduced $\chi^2$ of all fits to \emph{cycles} within a \emph{set} and the reduced $\chi^2$ of the Breit-Rabi fit are given in table~\ref{tab:DAQasdf}.
On the level of the achieved statistical precision no significant dependence of the 10 results on any of the changed experimental conditions could be found.
This justified to combine the 10 individual results into one weighted mean value.
Our final result deviates from the literature value by \mbox{\nuHF -- \nuHFlit =\unit[--3.4]{Hz}} with a total uncertainty of $\sigma_{\mathrm{tot}}$=\unit[3.8]{Hz}, which corresponds to a relative precision of  \unit[2.7]{ppb}.
The mean value is shown in Fig.~\ref{fig:big_plot}d as the dashed red line and the total  $1 \sigma$ uncertainties as the grey shaded area. 

The fit parameters $B_{\mathrm{osc}}$, $\bar{v}_H$ and $\sigma_v$,  which were fixed to a common average value for each set, were varied in order to assess the potential systematic uncertainties originating from the fit procedure.
The complete analysis was repeated 6 times with setting each of the three parameters individually to its lower and upper $1 \sigma$ boundary. The observed shifts of \nuHF\ for each parameter are listed in table \ref{tab:errorbudget}.
However, those three values added in quadrature yielded \unit[0.06]{Hz} and present a negligible systematic uncertainty.
The rubidium clock, which served as frequency standard, supplied a \unit[10]{MHz} reference signal to the microwave synthesiser.
A calibration was performed and revealed a shift of \unit[11.4]{mHz} or equivalently \unit[1.14]{ppb}.
This corresponds to \unit[1.6]{Hz} for $\nu_\text{HF}$ and is conservatively used as a $1 \sigma$ systematic uncertainty.
Table \ref{tab:errorbudget} summarises the error budget.

\begin{table}[h!]
\captionsetup{skip=0pt}
\caption{
%xxx\internallinenumbers
Error budget
}
\label{tab:errorbudget}
\begin{center}
\begin{tabular}{l c c c c c}
\hspace{0.5cm}  contribution & \multicolumn{3}{c}{1$\sigma$ st.dev. (Hz)}  \\
\hline
\hspace{0.5cm} systematic error & & &  \\
\hspace{1.0cm} frequency standard & & & 1.62  \\
\hspace{1.0cm} common fit parameters & & &   \\
\hspace{1.5cm} $\bar{v}_H$      & & & 0.05  \\
\hspace{1.5cm} $\sigma_v$       & & & 0.03  \\
\hspace{1.5cm} $B_\text{osc}$  & & & 0.02  \\
\hline
\hspace{0.5cm} systematic error total & & & 1.62  \\
\hspace{0.5cm} statistical error & & & 3.43 \\
\hline
\hline
total error & & & 3.79  \\
\end{tabular}
\end{center}
\end{table}

\section{Methods}

\begin{table*}[t]
\caption{
\textbf{Parameters of the data sets.}
Comparison of the 10 \emph{sets}. The 4 blocks of rows summarize (i) experimental conditions, (ii) statistics of the data acquisition, (iii) average fit parameter of \emph{cycles} and the average reduced chi-squares from applying fit-formula~(\ref{eq:lineshapeR}), and finally (iv)  the fit parameters and the corresponding reduced chi-square from applying the Breit-Rabi fit~(\ref{eq:Rabiformulafit}).
}
\label{tab:DAQasdf}
\begin{center}
\begin{tabular}{|p{0.12\textwidth}||p{0.07\textwidth}|p{0.07\textwidth}|p{0.07\textwidth}|p{0.07\textwidth}|p{0.07\textwidth}|p{0.07\textwidth}|p{0.07\textwidth}|p{0.07\textwidth}|p{0.07\textwidth}|p{0.07\textwidth}|}
\hline
 \emph{set} & 1 & 2 & 3 & 4 & 5 & 6 & 7 & 8 & 9 & 10 \\
\hline
%%% BLOCK (i) experimental conditions
\hline
PTFE tubing & straight & straight & straight & 90 deg. & 90 deg. & 90 deg. & 90 deg. & 90 deg. & 90 deg. & 90 deg. \\
\hline
cryostat temperature (K) & 23    & 16    & 100   & 50    & 50    & 50    & 50    & 50    & 50    & 50 \\
\hline
$d_\text{s}$ (mm) & 115   & 35    & 91    & 21    & 21    & 16    & 16    & 16    & 16    & 115 \\
\hline
cavity & \# 1     & \# 1     & \# 1     & \# 1     & \# 1     & \# 1     &\#  2     & \# 2     & \# 2     & \# 2 \\
\hline
precise monitoring of $I_{HC}$  & no    & no    & no    & yes   & yes   & yes   & yes   & yes   & yes   & yes \\
\hline
supercond. sextupole (A) & 350   & 350   & 400   & 350   & 350   & 350   & 350   & 350   & 350   & 350 \\
\hline
beam diameter (mm) & 8     & 8     & 8     & 8     & 8     & 8     & 8     & 8     & 22    & 22 \\
\hline
shielding layers & 2     & 2     & 2     & 2     & 2     & 2     & 1     & 2     & 2     & 2 \\
\hline
$I_{HC}$ polarity &  $\pm$     &   $\pm$    &    $\pm$   &     $+$  &      $\pm$ &      $\pm$ &     $\pm$  &    $\pm$   &   $\pm$    & $\pm$  \\
\hline
%%% BLOCK (iI) statistics of the data acquisition
\hline
number of \emph{scans}  & 8     & 6     & 6     & 10    & 12    & 12    & 16    & 16    & 12    & 12 \\
\hline
number of \emph{cycles}  & 23    & 46    & 26    & 50    & 60    & 60    & 80    & 80    & 60    & 60 \\
\hline
frequency data points  & 41    & 21    & 26    & 39    & 39    & 39    & 39    & 39    & 39    & 39 \\
\hline
acqu. time / data point (s) & 60    & 40    & 40    & 5     & 5     & 5     & 5     & 5     & 5     & 5 \\
\hline
%%% BLOCK (iii) fit parameter of cycles
\hline
$v$	$\newline$(m/s)	&	$1066\newline (1)$	&	$962\newline (2)$	&	$1152\newline (2)$	&	$888\newline (2)$	&	$857\newline (3)$	&	$883\newline (2)$	&	$933\newline (2)$	&	$922\newline (1)$	&	$1049\newline (1)$	&	$1131\newline (1)$	\\
\hline																						
$\sigma_\text{v}$	$\newline$(m/s)	&	$152\newline (2)$	&	$145\newline (3)$	&	$156\newline (2)$	&	$160\newline (2)$	&	$184\newline (2)$	&	$139\newline (2)$	&	$124\newline (2)$	&	$129\newline (2)$	&	$183\newline (1)$	&	$149\newline (1)$	\\
\hline
$B_{\text{osc}} $\newline$(10^{-7}$ T)  & $6.86\newline (0.01)$  & $6.49\newline (0.01)$  & $8.14\newline (0.01)$  & $5.73\newline (0.01)$  & $5.81\newline (0.01)$  & $5.78\newline (0.01)$  & $6.70\newline (0.03)$  & $6.28\newline (0.05)$  & $6.54\newline (0.03)$  & $6.93\newline (0.03)$ \\
\hline
$R_0$ \newline (Hz) & $27088\newline (232)$ & $24420\newline (576)$ & $26517\newline (234)$ & $26998\newline (458)$ & $20889\newline (237)$ & $26118\newline (84)$ & $23100\newline (194)$ & $24825\newline (225)$ & $56390\newline (2806)$ & $31584\newline (1724)$ \\
\hline
$\Delta R$ \newline(Hz) & $891\newline (30)$   & $476\newline (43)$   & $1112\newline (29)$  & $1471\newline (71)$  & $795\newline (47)$   & $1484\newline (50)$  & $1126\newline (42)$  & $1401\newline (51)$  & $4499\newline (907)$  & $3284\newline (154)$ \\
\hline
av. $\chi^2/n.d.f.$ of res. curves  & $2.6\newline (0.7)$   & $1.6\newline (0.5)$   & $1.9\newline (0.5)$   & $1.2\newline (0.3)$   & $1.0\newline (0.3)$   & $1.2\newline (0.2)$   & $1.1\newline (0.3)$   & $1.1\newline (0.2)$   & $1.9\newline (0.4)$   & $1.9\newline (0.4)$ \\
\hline

\hline
$B_{\text{res}} $\newline$(10^{-7}$ T)    & $4.0\newline (1.0)$   & $3.9\newline (2.6)$   & $2.3\newline (2.0)$   & $11.9\newline (19.1)$  & $5.4\newline (1.9)$   & $5.7\newline (1.0)$   & $3.1\newline (1.3)$  & $3.5\newline (1.1)$  & $2.5\newline (0.5)$  & $2.7\newline (0.5)$ \\
\hline
k$\newline$$(10^{-5}$ T/A)     & $45.83\newline (0.02)$ & $45.47\newline (0.16)$ & $45.70\newline (0.14)$ & $45.87\newline (0.37)$ & $45.76\newline (0.06)$ & $45.80\newline (0.03)$ & $45.85\newline (0.04)$ & $45.90\newline (0.03)$ & $45.90\newline (0.02)$ & $45.89\newline (0.02)$ \\
\hline
$\chi^2/n.d.f.$ of Breit-Rabi fit & 17.5/21  & 38.6/44  & 41.3/24  & 70.0/47  & 65.3/57  & 48.5/57  & 101.0/77& 83.2/77  & 75.0/57 & 55.6/57 \\
\hline
\nuHF\ - \nuHFlit\ (Hz) & $-12.0 \newline (10.6)$ & $22.6\newline (22.2)$  & $20.8\newline (19.4)$  & $19.7\newline (46.7)$  & $5.7\newline (23.9)$   & $-9.2\newline (12.8)$  &$ 7.1\newline (11.4)$   & $-2.9\newline (9.2)$  & $-8.8\newline (6.9)$  & $-3.5\newline (6.4)$ \\
\hline
\end{tabular}
\end{center}
\end{table*}

\subsection{Resonance line shape}

\begin{figure*}[htbp]
\begin{center}
\includegraphics[width=0.7 \textwidth]{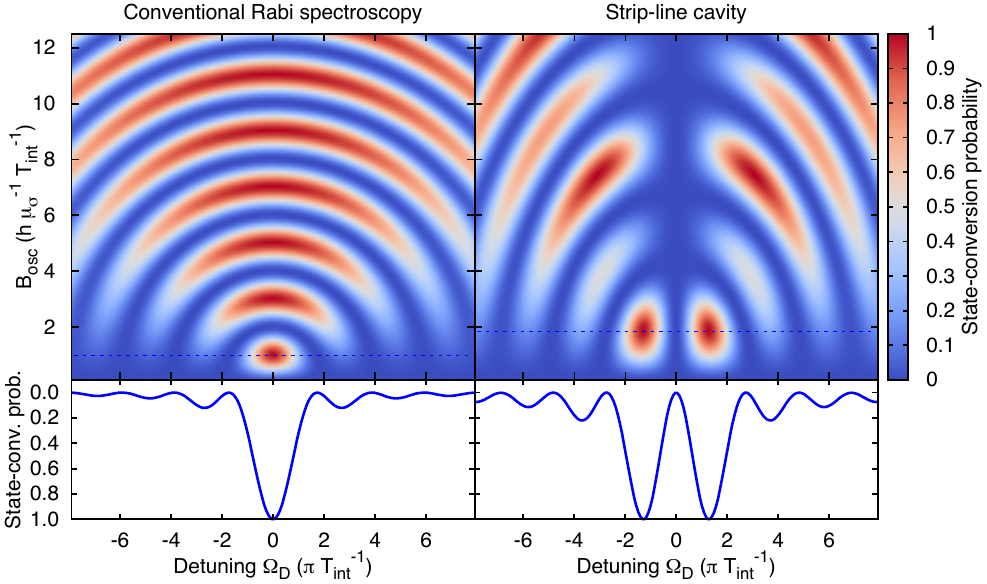}
\caption{
%xxx\internallinenumbers
\textbf{State-conversion probability maps for the detuning and driving strength.}
Comparison of the state-conversion probabilities as a function of the detuning $\Omega_\text{D}$ (in units of $\pi T_\text{int}^{-1}$) and the amplitude of the oscillating magnetic field $B_\text{osc}$ (in units of $h \mu_{+}^{-1} T_\text{int}^{-1}$) for the case of conventional Rabi spectroscopy (left) and when using a strip-line cavity (right) to drive the transition.
Both cases refer to a mono-energetic beam, which translates to a fixed interaction time $T_\text{int}$.
The dashed horizontal line indicates the required driving strength to reach the first complete state conversion.
The plots below are projections of the state-conversion probabilities at the dashed lines and show the ideal (i.e. mono-energetic) line shapes observed as count rate drops in the Rabi experiments.
}
\label{fig:powerdetuning_plot_comparison_RabiOurs}
\end{center}
\end{figure*}

The $\sigma_1$-transition in ground-state hydrogen is driven by an external microwave field, which is generated in a strip-line cavity and takes the form

\begin{equation}
\begin{split}
\mathbf{{B}_{\text{osc}}}(t) = B_{\text{osc}} \mathbf{e}_B \cos(\omega t)  \cos(\omega_{\text{cav}} t),  \\
\omega_{\text{cav}} =  \frac{\pi}{T_{\text{int}}} = \frac{\pi v_H}{L_{\text{cav}}} ,\\
0 < t < T_{\text{int}},
\end{split}
\end{equation}

where $\mathbf{e}_B$ is the unit vector pointing in the direction of the magnetic field ($z$-axis in the frame of the atoms, $x$-axis in the coordinate system of the experiment) and  $\nu=\omega/2 \pi$ is the applied microwave frequency.
The term $\cos(\omega_{cav} t)$ describes the changing amplitude of the magnetic field in the cavity along the beam propagation direction.
$T_\text{int}$ is the interaction time, which in turn follows from the hydrogen beam velocity $v_H$ and the length of the cavity $L_{\text{cav}}$.

The small external magnetic field is aligned parallel to the oscillating magnetic field, which only for the $\sigma_1$-transition leads to non-vanishing matrix elements. In addition the Zeeman shift separates the ground-state hydrogen sub levels by more than the observed resonance width.
Therefore, the transition dynamics is well described within the framework of the two-level system

\begin{equation}
\begin{split}
\label{eq:2LevelSystem}
| \phi \rangle = c_1(t) | \phi_1 \rangle + c_2(t) | \phi_2 \rangle , \\
| \phi_1 \rangle = |F=0,M_F=0 \rangle = \sqrt{\nicefrac{1}{2}} \left( | \uparrow^e \downarrow_p \rangle - |  \downarrow_e \uparrow^p \rangle \right) , \\
| \phi_2 \rangle = |F=1,M_F=0 \rangle = \sqrt{\nicefrac{1}{2}} \left( | \uparrow^e \downarrow_p \rangle + |  \downarrow_e \uparrow^p \rangle \right) , \\
\hat{\mathcal{H}}_{\text{atom}}  | \phi_i \rangle = E_i  | \phi_i \rangle , \\
|c_1(t)|^2 + |c_2(t)|^2 =1 .
\end{split}
\end{equation}

To obtain the time evolution of this system under the influence of the oscillating magnetic field the Hamiltonian needs to be extended by the interaction $\hat{\mathcal{H}}_{\text{int}} = -\hat{\pmb{\mu}} \cdot \mathbf{{B}_{osc}}(t)$ with

\begin{equation}
\label{eq:MagMomentOperator}
\hat{\pmb{\mu}} = - |g_e| \mu_B \frac{1}{\hbar} \mathbf{\hat{S}^e} + g_p \mu_N \frac{1}{\hbar} \mathbf{\hat{S}^p} ,
\end{equation}

for hydrogen.
Here, $\mathbf{\hat{S}^i}$ are the spin operators acting on the electron or proton spinor as indicated by the superscript.
An analytical solution can be found for conventional Rabi experiments, where the oscillating (or rotating) magnetic field has a constant amplitude $B_{osc}$ and doesn't include the term $\cos(\omega_{cav} t)$.
If the system is initially prepared purely in state $| \phi_1 \rangle$, then the conversion probability $|c_2|^2$ of finding it after a given interaction time $T_\text{int}$ in the second state $| \phi_2 \rangle$ depends on the strength of $B_{osc}$ and the detuning $\Omega_\text{D} = \omega-\omega_{12}$ with $\hbar \omega_{12} = E_2-E_1$:

\begin{equation}
|c_2|^2 = \frac{\Omega_{\text{R}}^2}{\Omega_\text{D}^2 + \Omega_{\text{R}}^2} \sin^2 \left( \frac{1}{2} \sqrt{\Omega_\text{D}^2 + \Omega_{\text{R}}^2} \ T_\text{int} \right) ,
\end{equation}

where $\Omega_{\text{R}}$ is the Rabi frequency, which is proportional to the amplitude of the oscillating magnetic field.
The relation for the $\sigma_1$-transition is

\begin{equation}
\Omega_{\text{R}} = \underbrace{ \left( |g_e| \mu_B + g_p \mu_N \right) }_{\mu_{+}} \frac{B_{\text{osc}}}{2 \hbar}.
\end{equation}

Including the term $\cos(\omega_{cav} t)$ requires numerical methods to determine the state-conversion probability.
Figure~\ref{fig:powerdetuning_plot_comparison_RabiOurs} shows a comparison of $|c_2|^2$ as a function of the detuning $\Omega_{\text{D}}$ and the driving strength $B_{\text{osc}}$ of conventional Rabi spectroscopy and the strip-line cavity designed for the antihydrogen experiment.
The latter case features the distinct \emph{double-dip} structure with vanishing effects at the actual transition frequency.
For a given interaction time $T_\text{int}$ the best precision is achieved with the first full state conversion in both situations.
For the conventional case this corresponds to a so-called $\pi$-pulse indicating that the condition $\Omega_{\text{R}} \cdot T_{\text{int}} = \pi$ is satisfied or alternatively $B_{\text{osc}}=h \mu_{+}^{-1} T_\text{int}^{-1}$.
The \emph{double-dip} resonance reaches the first full state conversion when applying a somewhat stronger oscillating magnetic field $B_{\text{osc}} \sim 1.86 \times h \mu_{+}^{-1} T_\text{int}^{-1}$.

\begin{figure*}[htbp]
\begin{center}
\includegraphics[width=0.7 \textwidth]{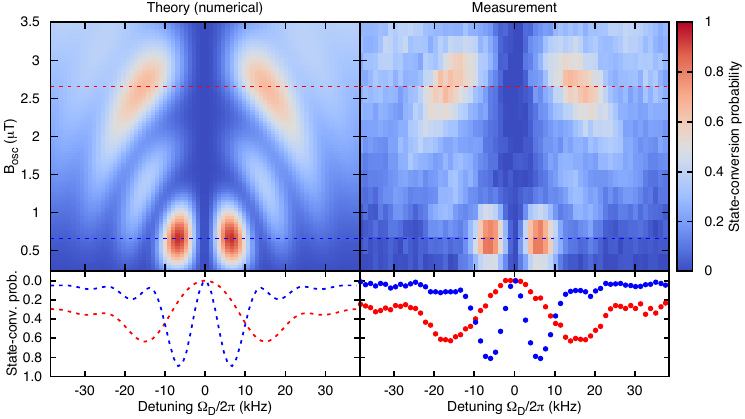}
\caption{
% %xxx\internallinenumbers
\textbf{State-conversion probability maps convoluted with a velocity distribution.}
Comparison of theoretical and measured state-conversion probabilities as a function of the detuning $\Omega_\text{D} / 2 \pi$ in units of kHz and the amplitude of the oscillating magnetic field $B_\text{osc}$ in units of $\mu$T. The theoretical map includes the effect of a Gaussian-like velocity distribution with $\bar{v}_H$=\unit[1060]{m/s} and $\sigma_v$=\unit[95]{m/s}. The measurement was taken setting a large distance between the permanent sextupole magnets of $d_\text{s}$=\unit[115]{mm} as it has been used for the \emph{sets} 1 and 10. The blue and red dashed horizontal lines indicate the driving strengths, where the first and second full state conversion would be reached in the case of a mono energetic beam. The plots below are projections of the state-conversion probabilities at the dashed lines showing good agreement between theory and measurement. Frequency spectra across the narrow \emph{double-dips} at the first state conversion yield the highest precision.
}
\label{fig:MapsvDistrTheoExp}
\end{center}
\end{figure*}

From a 2D-map as shown in Fig.~\ref{fig:powerdetuning_plot_comparison_RabiOurs} fit functions of the state-conversion probabilities for a mono-energetic beam can be derived with $\nu_c$, the strength of $B_{\text{osc}}$, and the hydrogen beam velocity $v_H$ as fit parameters.
This was realised by constructing a 2D-spline interpolation $\mathcal{S}(\nu;\nu_c,B_{\text{osc}},v_H)$ to the numerically generated state-conversion probabilities at discrete points.
A more realistic resonance line shape is then obtained by including the effect of the velocity distribution of the hydrogen beam, which translates to a not sharply defined interaction time $T_\text{int}$.
Note that both axis of the 2D-maps are normalised to $T_\text{int}^{-1}$.
Therefore, on an absolute scale for $\Omega_\text{D}$ and $B_{\text{osc}}$ a change of $T_\text{int}$ is equivalent to a 2D-zooming of the state-conversion probability map.
The roughly Gaussian velocity distribution of the hydrogen beam after passage of the polarising and velocity-selecting permanent sextupole magnets is approximated by binomial coefficients for a discrete numerical realisation of the convolution

\begin{equation}
\label{eq:lineshape}
\begin{split}
\mathcal{F}(\nu;\nu_c,B_{\text{osc}},\bar{v}_H,\sigma_v) = \\
= 2^{-N} \sum_{n=0}^{N} \binom{N}{n} \mathcal{S}(\nu; \nu_c, B_{\text{osc}}, v_{(N,n)}) , \\
v_{(N,n)} = \bar{v}_H+(n-N/2) dv, \\
dv \sim 2 \sigma_v N^{-1/2}.
\end{split}
\end{equation}

The result of a convolution with such a velocity distribution is illustrated in Fig.~\ref{fig:MapsvDistrTheoExp} and compared to a measured map.
The present analysis used $N=6$ because choices of $N>6$ did not change nor improve the fit results.
For completeness two more fit parameters were needed.
In order to scale the state-conversion probability, which is a number between 0 and 1, to the observed count rates, a count rate baseline $R_0$ and a count rate drop for complete state conversion $\Delta R$ were introduced

\begin{equation}
\label{eq:lineshapeR}
\begin{split}
\mathcal{F}_{R}(\nu;\nu_c,B_{\text{osc}},\bar{v}_H,\sigma_v,R_0,\Delta R) = \\
= R_0 - \Delta R \cdot \mathcal{F}(\nu;\nu_c,B_{\text{osc}},\bar{v}_H,\sigma_v),
\end{split}
\end{equation}

which are albeit not relevant for the obtained results.

\section{Acknowledgement}

We want to thank H.~Kundsen, H.-P.~E.~Kristiansen, F.~Caspers, T.~Kroyer, S.~Federmann, P.~Caradonna, M.~Wolf, M.~Heil, F.~Pitters, C.~Klaushofer, S.~Friedreich, and B.~W\"unschek for their contributions. We acknowledge technical support by the CERN Cryolab and Instrumentation group TE-CRG-CI as well as the CERN Magnet Normal Conducting group TE-MSC-MNC.

This work has been supported by
the European Research Council under European Union's Seventh Framework Programme (FP7/2007-2013)/ ERC Grant agreement (291242),
the Austrian Ministry of Science and Research,
the Austrian Science Fund (FWF): W1252-N27.

\section{Author Contributions}

M.D.	prepared and performed the experiment, carried out the data analysis, and wrote parts of the initial manuscript.
C.B.J. performed the experiment and carried out parts of the data analysis.
B.K. and C.S. performed simulations.
C.M.	and M.C.S. prepared and performed the experiment, guided and carried out parts of the data analysis, and wrote the manuscript.
O.M. and J.Z. prepared the experiment.
E.W.	proposed and prepared the experiment, guided the data analysis, and wrote the manuscript.

\end{document}